# Top-gated graphene field-effect-transistors formed by decomposition of SiC

Y. Q. Wu, P. D. Ye [a)], M.A. Capano [b)], Y. Xuan, Y. Sui, M. Qi, and J.A. Cooper

*School of Electrical and Computer Engineering and Birck Nanotechnology Center,
Purdue University, West Lafayette, IN 47907*

T. Shen, D. Pandey, G. Prakash, and R. Reifenberger

*Department of Physics and Birck Nanotechnology Center,
Purdue University, West Lafayette, IN 47907*

Top-gated, few-layer graphene field-effect transistors (FETs) fabricated on thermally-decomposed semi-insulating 4H-SiC substrates are demonstrated. Physical vapor deposited $SiO_2$ is used as the gate dielectric. A two-dimensional hexagonal arrangement of carbon atoms with the correct lattice vectors, observed by high-resolution scanning tunneling microscopy, confirms the formation of multiple graphene layers on top of the SiC substrates. The observation of n-type and p-type transition further verifies Dirac Fermions' unique transport properties in graphene layers. The measured electron and hole mobility on these fabricated graphene FETs are as high as 5400 $cm^2$/Vs and 4400 $cm^2$/Vs respectively, which are much larger than the corresponding values from conventional SiC or silicon.

[a)]Electronic mail: yep@purdue.edu
[b)]Electronic mail: capano@purdue.edu

To a large extent, the success of modern microelectronics is based on the continuous miniaturization or scaling of silicon metal-oxide-semiconductor field-effect transistors (MOSFET), which makes them smaller, faster, and cheaper. However, silicon-based technology will encounter physical and technical limits within the next decade, which motivates the semiconductor industry to explore alternative device technologies such as Ge[1], III-Vs[2], and carbon nanotubes (CNTs) [3-5] to replace silicon as active channel materials. Graphene, a monolayer of carbon atoms tightly packed into a two-dimensional (2D) hexagonal lattice, has recently been shown to be thermodynamically stable and exhibit astonishing transport properties, such as an electron mobility of ~15,000 $cm^2$/Vs and velocity of ~$10^8$ cm/s [6]. High-quality monolayer graphene has been obtained in small (few micron) areas by exfoliation of highly-ordered pyrolytic graphite (HOPG) and transferred onto $SiO_2$ substrates for further device fabrication [7,8]. However, this exfoliation process cannot form the basis for a large-scale manufacturing process. Recent reports of large-area epitaxial graphene by thermal decomposition of SiC wafers have provided the missing pathway to a viable electronics technology [9]. The advantage of epitaxially-grown graphene for nanoelectronic application resides in its planar 2D structure that enables conventional top-down lithography and processing techniques.

Three-terminal MOSFETs constitute the backbone of today's silicon industry, and similar device structures are envisioned for graphene. For these devices to be successful, carrier transport in the conduction channel must be controlled efficiently by the top-gate structure. Some attempts to address these issues were reported very recently, both on peeled graphene flakes [10] or SiC surfaces after graphitization [11]. In this letter, the observation of n-type and p-type transition on epitaxially-grown graphene films by top-gate bias is reported. More importantly, the measured electron and hole mobilities on fabricated top-gate graphene field-effect transistors exceeds 5400 cm$^2$/Vs and 4400 cm$^2$/Vs, respectively. This is one order-of-magnitude higher than earlier reported [11], and approaches the values obtained from exfoliated graphene films [6-8].

The graphene films were grown on the carbon or silicon face of semi-insulating 4H-SiC substrates in an Epigress VP508 SiC hot-wall chemical vapor deposition (CVD) reactor. The off-cut angle of the substrate is nominally zero degrees. Prior to growth, substrates are subjected to a hydrogen etch at 1600 °C for 5 minutes, followed by cooling the samples to below 700 °C. After evacuating hydrogen from the system, the growth environment is pumped to an approximate pressure of 2×10$^{-7}$ mbar before temperature ramping at a rate of 10-20 °C/min. up to a specified growth temperature. Growth temperatures investigated ranged from 1350 °C – 1650 °C. Scanning tunneling microscopy (STM) images taken at a constant set point current of 0.5 nA with a bias of -0.5 V revealed nanometer high steps as illustrated in Figure 1(a), implying that graphene follows the 4H-SiC surface topography after H$_2$ etching. The surface roughness of a plateau was measured to be ~0.2 nm. This morphology is critical in achieving high-mobility graphene films on SiC. Atomic resolved STM images within a graphene-like domain were performed in constant height mode. Figure 1(b) shows a typical atomic-resolution STM image after denoising using a FFT filter. The image is superimposed with a triangular lattice with lattice vectors of 0.25 nm, agreeing well with the expected 0.246 nm lattice spacing of HOPG. This is far different from the 0.307 nm lattice constant of 4H SiC. The observed triangular lattice is expected from the Bernal stacking of neighboring graphene sheets, as in HOPG.[12] The formation of multiple graphene layers on SiC after high temperature decomposition is further confirmed by the similar measurement on HOPG, where a similar triangular sublattice is observed with the lattice vectors of 0.26 nm.

The device structure of the fabricated graphene FET is shown in the inset of Fig. 2(a). Device isolation of the graphene film was realized by 120 nm deep SF$_6$ based dry etching with photo-lithographically defined photoresist as a protection layer. Ti/Au metallization was used to form Ohmic contacts on graphene as source and drain. Physical vapor deposition (PVD) SiO$_2$ was used as top-gate dielectric. Plasma-based processing was avoided to prevent damage to the graphene films. Chemical vapor deposition (CVD) or atomic layer deposition (ALD) cannot form high-quality dielectrics on the basal plane of graphene without a non-covalent functionalization layer.[13-15] Finally, conventional Ni/Au metals were electron-beam evaporated, followed by lift-off to form the gate electrodes. The process requires three levels of lithography (isolation, ohmic, and gate), all done using a contact printer.

Figure. 2(a) shows the dc $I_d$−$V_{ds}$ characteristic with a gate bias from -1 to 2.5 V on a device with a fully-covered gate, as shown in the inset. The measured graphene FET has a designed gate length ($L_g$) of 400 μm and a gate width ($L_w$) of 50 μm. The gate oxide is 50 nm $SiO_2$. The gate leakage current is very low, below $10^{-9}$ A under the same bias conditions, corresponding to a gate leakage current density of $3\times10^{-6}$ $A/cm^2$. The drain current can be modulated by ~50% with a few volts gate bias. A reference SiC sample without the graphitization treatment, processed at the same time, shows no current (< 40 pA). Figure 2(b) illustrates the transfer characteristics of this graphene FET at $V_{ds}$ = 2 V. The drain current or conductance exhibits a region of minimum at $V_{gs}$ ~ -0.8 V. This dip in single-graphene layer is well documented, and corresponds to a minimum conductivity of ~ $4e^2/h$ at the charge neutrality point or Dirac point, where $e$ and $h$ are the electric charge and Planck's constant, respectively. It is the same physical origin here for multiple graphene films. The device cannot be turned off to zero drain current or conductance, since graphene or graphite is a semi-metal with no bandgap. A finite bandgap can be created by patterning graphene into graphene nano-ribbons,[16-17] but the dimensions of such ribbons press the limits of modern lithography. A finite bandgap may also be created by breaking the graphene sublattice symmetry using selective doping,[18] a vertical electric field,[19] or interactions with the underlying substrate.[20] With the conductance of $1.25\times10^{-4}$ S in this particular device, we estimate that this few-layer graphene film could contain 6-7 layers of graphene. The slope of the drain current shows that the peak extrinsic transconductance ($G_m$) is ~ 1.4 mS/mm, due to its extraordinarily large gate length. Assuming that epitaxially grown few-layer graphene on SiC has the same layer-by-layer separation of 0.335 nm as HOPG, the conduction channel of this particular device should be constrained within ~2.0 nm. The channel mobility can be simply estimated by $\mu=[(\Delta I_d/V_{ds})/(L_w/L_g)]/C_{ox}\Delta V_{gs}$ using the surface-channel device formula. Here, $C_{ox}$ is determined by $\varepsilon_0\varepsilon_r A/d$, where $\varepsilon_0$ is permittivity of free space, $\varepsilon_r$ is 3.0 for PVD $SiO_2$ without annealing, A is unit area, and d is the gate oxide thickness. The extracted electron effective mobility is as high as 5,400 $cm^2$/Vs, one order-of-magnitude higher than the value reported in Ref. [11], and approaching values reported in exfoliated graphene films.[6-8] In contrast to Ref. [10], no significant degradation of mobility is observed due to the use of PVD $SiO_2$. The device characteristics could become ambipolar like at low charge density or near Dirac point [14,21,22]. A simple approach, proposed by Huard et al., is used to further characterize the sample by extracting the p-n junction resistance[21], as shown by the dashed lines in Figure 2(b). The corrected electron mobility drops 10-30% around Dirac point $V_{gs}$=-0.8V. Note that this correction does not have to be needed at unipolar condition or at high charge density.

To accurately control the number of graphene layers formed by decomposition of the SiC surface remains a challenge. Figure 3(a) shows graphene sheet resistivity as a function of growth temperature for ten growth runs. It is obvious that the number of graphene layers can be controlled by decomposition temperature, time, and other processing conditions. 1500 -1550 °C is the growth temperature window for monolayer graphene films on Si face of SiC in the CVD system used in these experiments. Figure 3(b) illustrates drain conductance vs. $V_{gs}$ characteristics as a function of $V_{ds}$ on a similar graphene FET with the possibility of only one monolayer. The drain current at low drain voltage $V_{ds}$ = 0.2V is around $2\times10^{-6}$ A, which corresponds to a sheet conductivity of 2 $e^2/h$, the same order as 4 $e^2/h$ for a single graphene layer. A clear n-type to p-type transition, a critical feature for the graphene band structure, is also observed after gate bias sweeps beyond -20 V. The $SiO_2$ is 120 nm thick, with the gate leakage current less than $10^{-7}$ A under gate bias from -22 V to 20 V. From the simple calculation described above, the hole mobility and electron mobility are 4,400 cm$^2$/Vs and 2,000 cm$^2$/Vs, respectively. The origin of the observed flat features at the n-type to p-type transition is unclear. It could be introduced by substrate leakage current related to the relative low substrate resistivity (on the order of $10^5$ Ω cm) of the semi-insulating SiC substrate. This does not exclude the possibility of opening a bandgap in the top monolayer of graphene, coupled with a buffer graphene layer.[20] In particular, the narrowing plateau feature with increasing drain voltage fits the bandgap picture well. It could also be related to the co-existence of thermally-activated electrons and holes near the Dirac point at room temperature [23]. It could also be related with creating a large fluctuating region, where average carrier density is smaller than disorder or domain boundary induced fluctuations [24]. More experimental and theoretical studies are needed to clarify these points. The drain conductance is not independent of drain voltage, since device operation could become complicated if part of the channel is negatively gate-biased and part is positively gate-biased, as referenced to the neutrality point or Dirac point.

In summary, few-layer graphene FET fabricated on decomposed semi-insulating SiC substrates with SiO$_2$ as gate dielectric is fabricated. A clear n-type to p-type transition at the Dirac point is observed. A maximum electron mobility of 5,400 cm$^2$/Vs and hole mobility of 4,400 cm$^2$/Vs is achieved. Graphene is a promising electronic material for future nanoelectronics, due to its high carrier mobility and high Fermi velocity compared to conventional semiconductors.

The authors would like to thank J. Appenzeller, M. S. Lundstrom, and L. Rokhinson for the valuable discussions.

**Figure Captions**

Figure 1. The graphene layer was studied using an Omicron ultra-high-vacuum STM operating at room temperature in the low $10^{-10}$ Torr regime. Etched W tips were used throughout. The XY piezo calibration was checked independently by imaging atoms in HOPG. (a) STM image of a graphene film formed on a carbon-face on-axis semi-insulating 4H SiC substrate. The 1.2 nm or 1.4 nm steps are unit steps in the basal plane of the SiC substrate. (b) Atomic resolution STM image of a triangular sublattice of carbon atoms in a graphene film. Due to the offset between two sublattices in a multiple graphene film, only every other surface atom is imaged.

Figure 2. (a) $I_{ds}$ vs. $V_{ds}$ as a function of $V_{gs}$ on a few-layer graphene FET measured at room temperature. $V_{gs}$ changes from 2.5 V (top) to -1.0 V (bottom) in 0.5 V steps. (b) Transfer characteristics of the same device at $V_{ds}$ = 2 V. The dashed lines are corrected one around Dirac point after extracting the p-n junction resistance or "odd" resistance component (see Ref. 21).

Figure 3. (a) Sheet resistivity of graphene films formed on SiC as a function of growth temperature. The gray line is a guide to the eye. (b) Drain conductance vs. $V_{gs}$ as a function of $V_{ds}$ on a possible monolayer graphene FET measured at room temperature. The graphene film is formed on Si face of SiC. A clear n-type to p-type transition is obtained once the gate bias is strongly positive or negative.

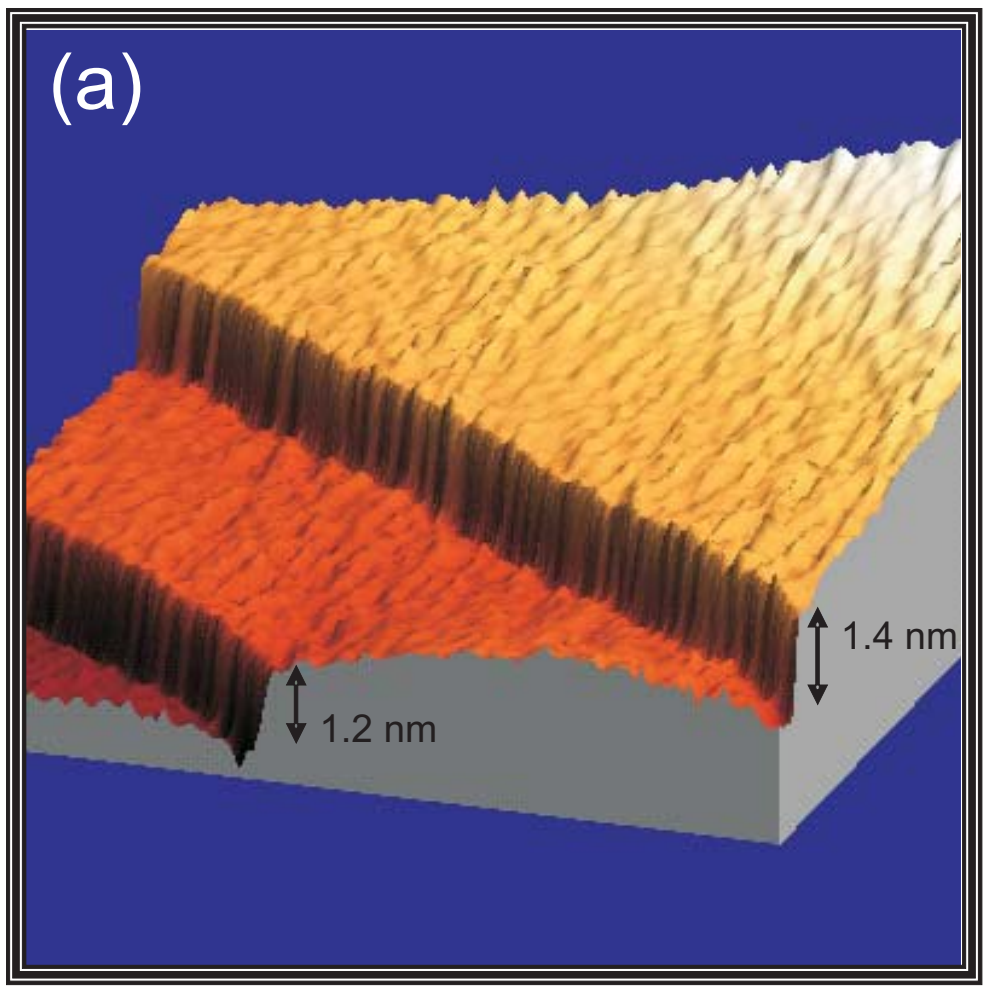 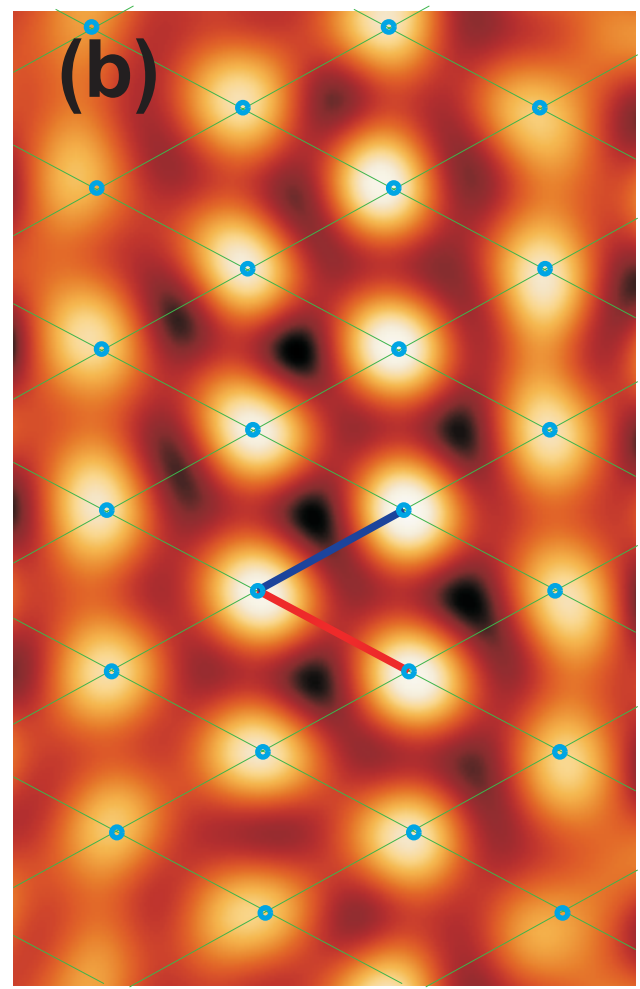

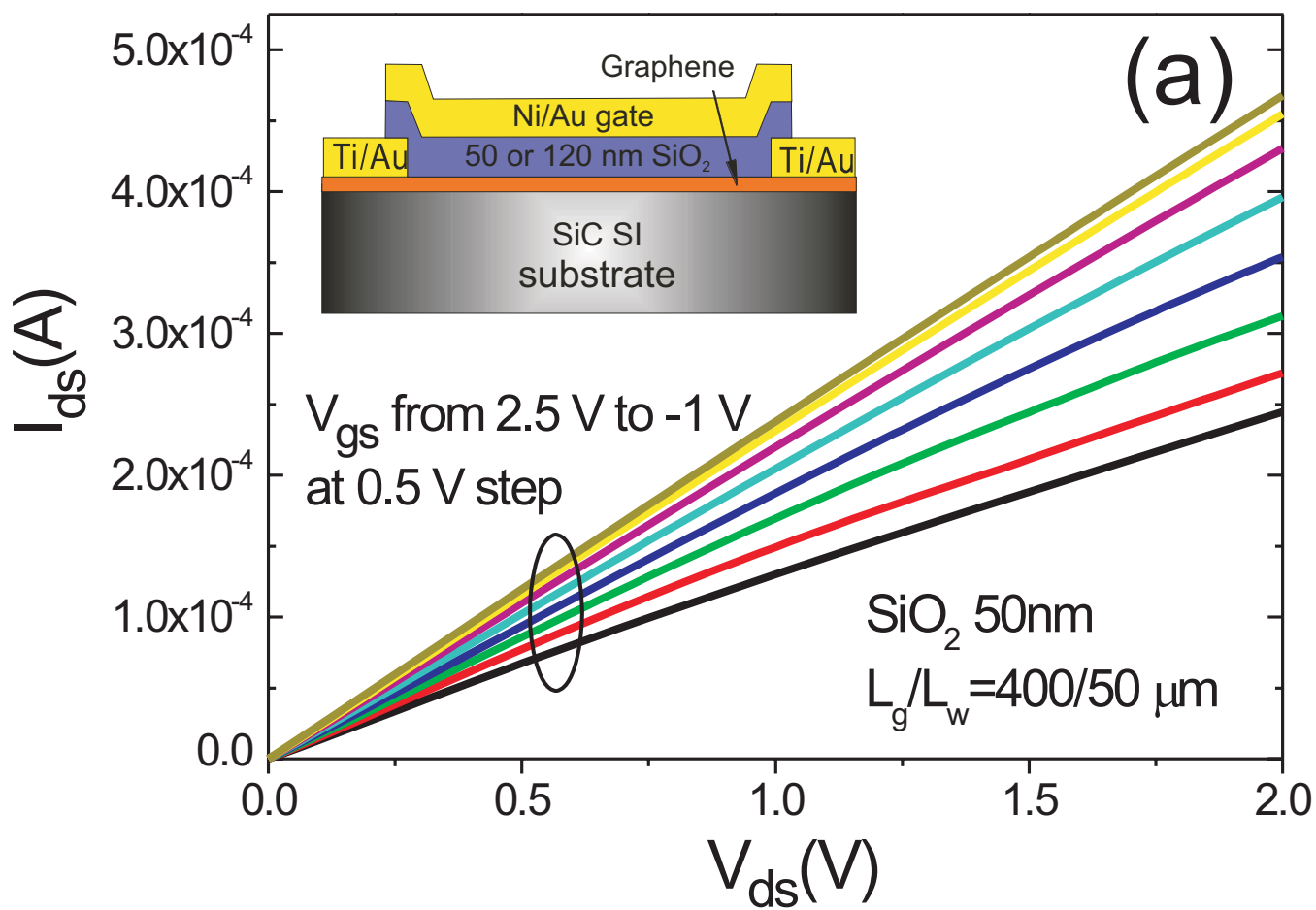

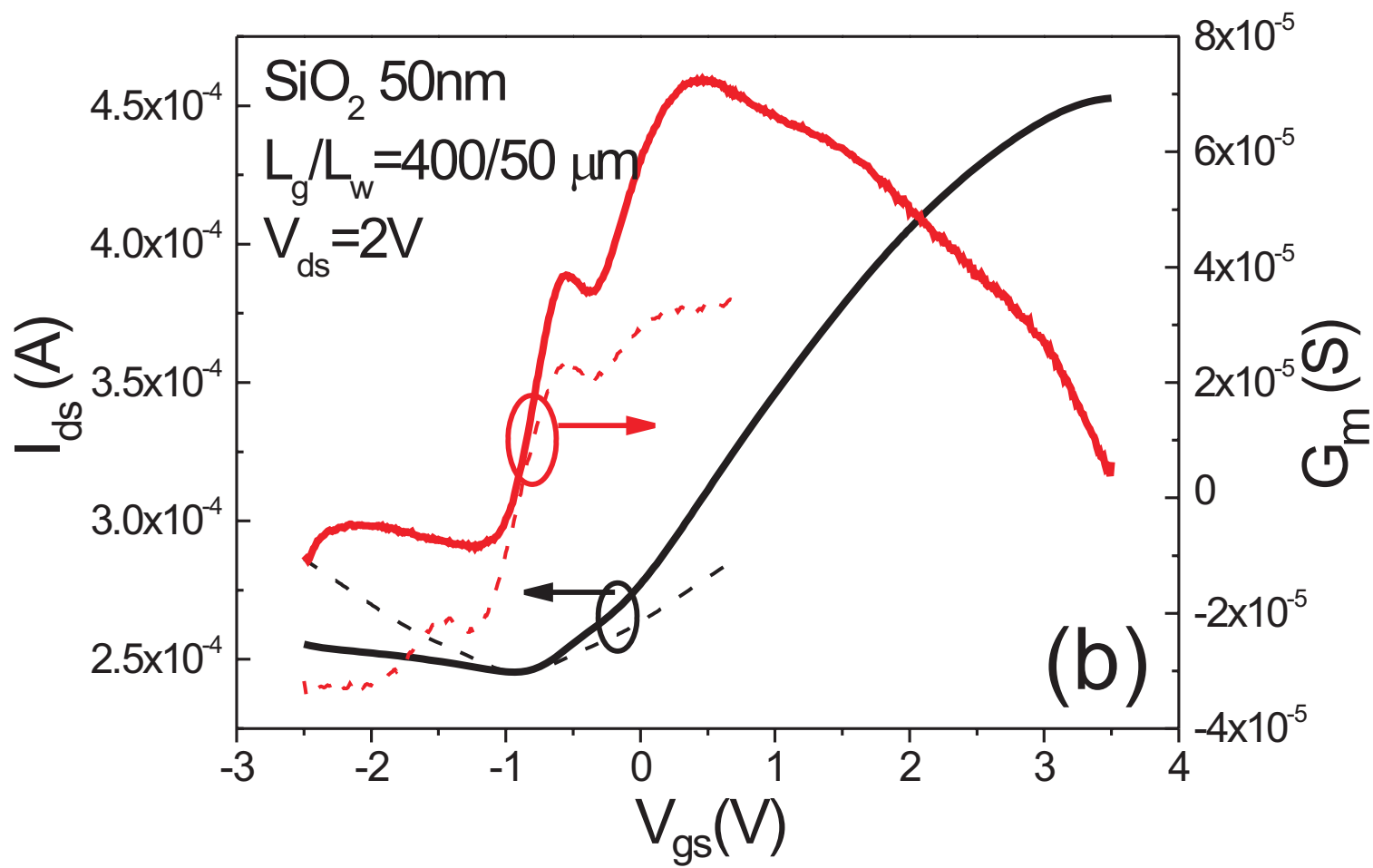

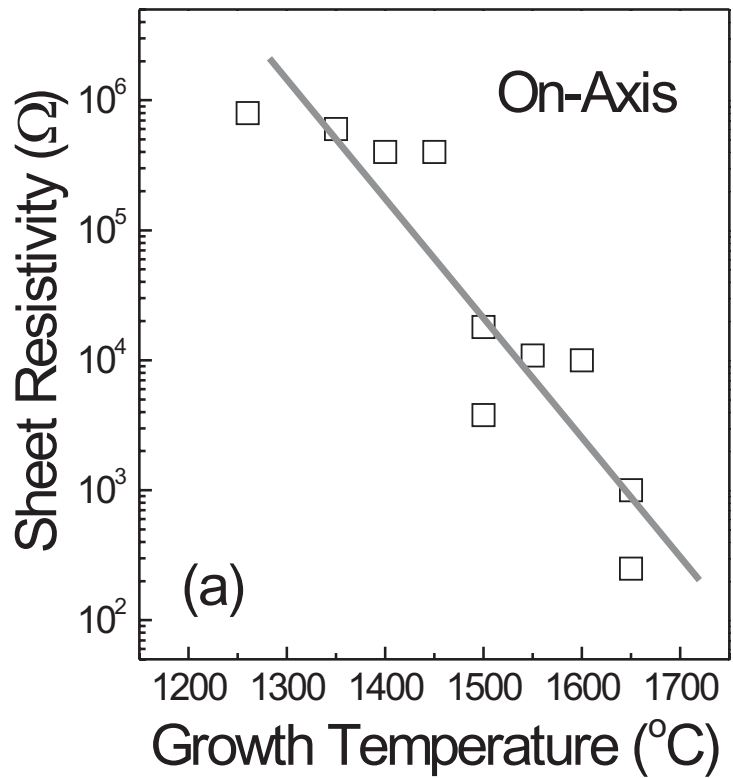 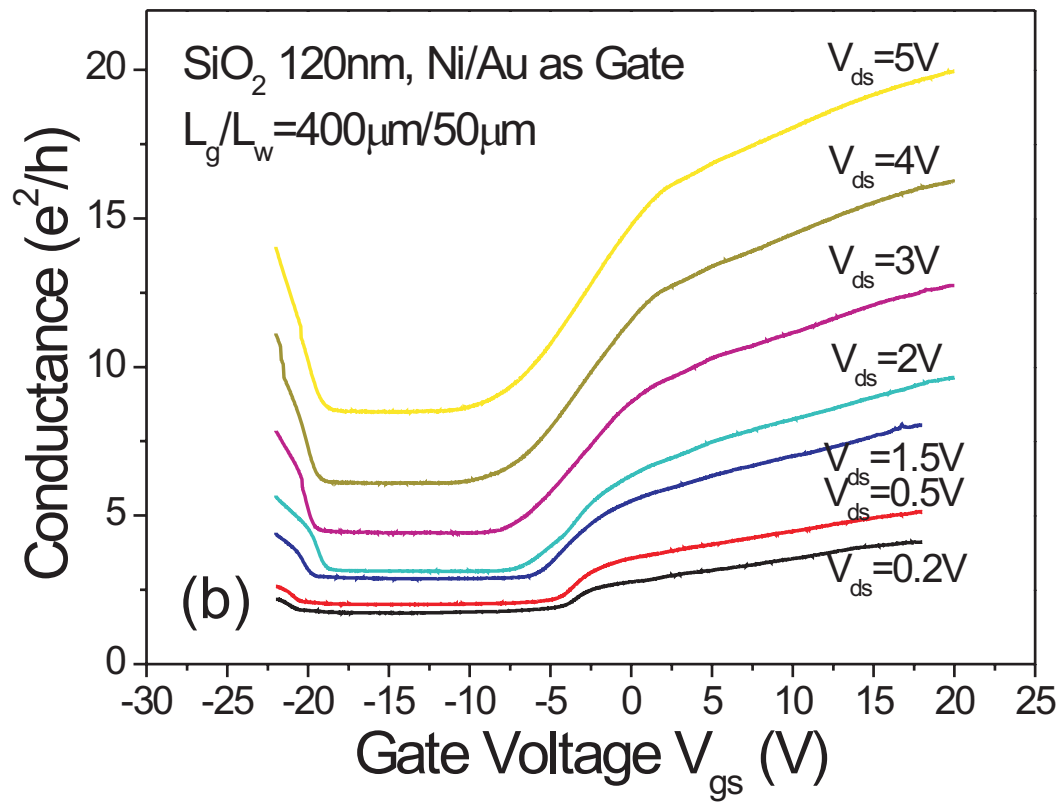